\documentstyle[12pt]{article}
\begin{document}

\title{\Large\bf Transformation of second-class into first-class
constraints in supersymmetric theories}

\author{J. Barcelos-Neto\thanks{\noindent e-mail:
barcelos@if.ufrj.br} ~and W. Oliveira\thanks{\noindent e-mail:
wilson@fisica.ufjf.br -- Permanent address: UFJF -  Departamento de
F\'{\i}sica - Juiz de Fora, MG 36036-330}\\
Instituto de F\'{\i}sica\\ 
Universidade Federal do Rio de Janeiro\\
RJ 21945-970 - Caixa Postal 68528 - Brasil\\
\date{}}

\maketitle
\abstract
We use the method due to Batalin, Fradkin, Fradkina, and Tyutin
(BFFT) in order to convert second-class into first-class constraints
for some quantum mechanics supersymmetric theories. The main point to
be considered is that the extended theory, where new auxiliary
variables are introduced, has to be supersymmetric too. This leads to
some additional restrictions with respect the conventional use of the
BFFT formalism.

\vfill
\vspace{1cm}
\newpage

\section{Introduction}

\bigskip
The method developed in series of papers by Batalin, Fradkin,
Fradkina, and Tyutin (BFFT) \cite{BFF,BT} has as main purpose the
transformation of second-class into first-class constraints
\cite{Dirac}. This is achieved with the aid of auxiliary fields that
extend the phase space in a convenient way. After that, one has a
gauge-invariant system which matches the original theory when the
so-called unitary gauge is chosen.

\medskip
The BFFT method is quite elegant and operates systematically. One
application we could immediately envisage for it is the covariant
quantization of systems with the Green-Schwarz (GS) supersymmetry
\cite{GS}. In these systems, the existence of the $\kappa$-symmetry
\cite{Siegel} inhibits a covariant separation of first and
second-class constraints. The use of the method there would transform
the second-class constraints into first-class and one could operate
with them in the conventional way.  However, this might not be an easy
and simple task as one may confirm by looking at the works of
covariant quantization of superstrings \cite{Kallosh}. We intend that
there are some problems that should be solved first. One of the
crucial question is how the BFFT method can operate consistently in
supersymmetric theories. This is a pertinent question because when
the phase space is extended by introducing auxiliary variables a new
theory is obtained, and this must also be supersymmetric.

\medskip
The purpose of the present paper is to address the BFFT method to
this particular problem related to supersymmetric theories. We show
that the freedom we have in choosing some parameters in the
application of the method can be conveniently used in order that a
final supersymmetric theory should be verified. We consider two
examples in supersymmetric quantum mechanics and we show that the
method is only consistently applied when an infinite number of
first-class constraints is introduced. 

\medskip
Our paper is organized as follows. In Sec. 2 we make a brief review
of the BFFT method in order to emphasize and clarify some of its
particularities that we shall use in the forthcoming sections. In
Sec. 3 we consider one of the simplest supersymmetric quantum
mechanical model, where the supersymmetric space is described by just
one time and one Grassmannian variable.  This system does not permit
us to introduce interaction and has to be analyzed with bosons and
fermions disconnected. In Sec. 4 we consider a more realistic model
where it is possible to have interactions between bosons and
fermions. This system is described in a space with one time and two
Grassmannian variables. Finally, Sec. 5 contains some concluding
remarks and we also introduce an appendix in order to explain some
details concerning the choice of the extended space.

\vspace{1cm}
\section{Brief review of the BFFT formalism}
\renewcommand{\theequation}{2.\arabic{equation}}
\setcounter{equation}{0}

\bigskip
Let us consider a system described by a Hamiltonian $H_0$ in a
phase-space $(q^i,p^i)$ with $i=1,\dots,N$. Here we suppose that the
coordinates are bosonic (extension to include fermionic degrees of
freedom and to the continuous case can be done in a straightforward
way). It is also supposed that there just exist second-class
constraints (at the end of this section we refer to the case where
first-class constraints are also present). Denoting them by $T_a$,
with $a=1,\dots ,M<2N$, we have

\begin{equation}
\label{2.1}
\bigl\{T_a,\,T_b\bigr\}=\Delta_{ab}\,,
\end{equation}

\bigskip\noindent
where $\det(\Delta_{ab})\not=0$. 

\medskip
The general purpose of the BFFT formalism is to convert second-class
constraints into first-class ones. This is achieved by introducing
canonical variables, one for each second-class constraint (the
connection between the number of second-class constraints and the new
variables in a one-to-one correlation is to keep the same number of
the physical degrees of freedom in the resulting extended theory). We
denote these auxiliary variables by $y^a$ and assume that they have
the following general structure

\begin{equation}
\label{2.2}
\bigl\{y^a,\,y^b\bigr\}=\omega^{ab}\,,
\end{equation}

\bigskip\noindent
where $\omega^{ab}$ is a constant quantity with
$\det\,(\omega^{ab})\not=0$. The obtainment of $\omega^{ab}$ is
discussed in what follows. It is embodied in the calculation of the
resulting first-class constraints that we denote by $\tilde T_a$. Of
course, these depend on the new variables $y^a$, namely

\begin{equation}
\label{2.3}
\tilde T_a=\tilde T_a(q,p;y)\,,
\end{equation}

\bigskip\noindent 
and satisfy the boundary condition

\begin{equation}
\label{2.4}
\tilde T_a(q,p;0)=\tilde T_a(q,p)\,.
\end{equation}

\bigskip\noindent
Another characteristic of these new constraints is that they are
assumed to be strongly involutive, i.e.

\begin{equation}
\label{2.5}
\bigl\{\tilde T_a,\,\tilde T_b\bigr\}=0\,.
\end{equation}

\bigskip
The solution of~(\ref{2.5}) can be achieved by considering $\tilde T_a$
expanded as

\begin{equation}
\label{2.6}
\tilde T_a=\sum_{n=0}^\infty T_a^{(n)}\,,
\end{equation}

\bigskip\noindent
where $T_a^{(n)}$ is a term of order $n$ in $y$. Compatibility
with the boundary condition~(\ref{2.4}) requires that

\begin{equation}
\label{2.7}
T_a^{(0)}=T_a\,.
\end{equation}

\bigskip\noindent
The replacement of~(\ref{2.6}) into~(\ref{2.5}) leads to a set of
equations, one for each coefficient of $y^n$. We list some of them
below

\begin{eqnarray}
\bigl\{T_a^{(0)},T_b^{(0)}\bigr\}_{(q,p)}
&\!\!\!\!+\!\!\!\!&\bigl\{T_a^{(1)},T_b^{(1)}\bigr\}_{(y)}=0\,,
\label{2.8a}\\
\bigl\{T_a^{(0)},T_b^{(1)}\bigr\}_{(q,p)}
&\!\!\!\!+\!\!\!\!&\bigl\{T_a^{(1)},T_b^{(0)}\bigr\}_{(q,p)}
+\bigl\{T_a^{(1)},T_b^{(2)}\bigr\}_{(y)}
+\bigl\{T_a^{(2)},T_b^{(1)}\bigr\}_{(y)}=0\,,
\label{2.8b}\\
\bigl\{T_a^{(0)},T_b^{(2)}\bigr\}_{(q,p)}
&\!\!\!\!+\!\!\!\!&\bigl\{T_a^{(1)},T_b^{(1)}\bigr\}_{(q,p)}
+\bigl\{T_a^{(2)},T_b^{(0)}\bigr\}_{(q,p)}
+\bigl\{T_a^{(1)},T_b^{(3)}\bigr\}_{(y)}
\nonumber\\
&\!\!\!\!+\!\!\!\!&\bigl\{T_a^{(2)},T_b^{(2)}\bigr\}_{(y)}
+\bigl\{T_a^{(3)},T_b^{(1)}\bigr\}_{(y)}=0\,,
\label{2.8c}\\
&\vdots&
\nonumber
\end{eqnarray}

\bigskip\noindent 
These correspond to coefficients of the powers $y^0$, $y^1$,
$y^2$, $\dots$, respectively. The notation used above,
$\{,\}_{(q,p)}$ and $\{,\}_{(y)}$, represents the parts of the
Poisson bracket $\{,\}$ relative to the variables $(q,p)$ and
$(y)$.

\medskip
Equations above are used iteratively in the obtainment of the
corrections $T^{(n)}$ ($n\geq1$). Equation~(\ref{2.8a}) shall give
$T^{(1)}$. With this result and~(\ref{2.8b}), one calculates
$T^{(2)}$, and so on. Since $T^{(1)}$ is linear in $y$ we may
write

\begin{equation}
\label{2.9}
T_a^{(1)}=X_{ab}(q,p)\,y^b\,.
\end{equation}

\bigskip\noindent 
Introducing this expression into~(\ref{2.8a}) and using the boundary
condition~(\ref{2.4}), as well as~(\ref{2.1}) and~(\ref{2.2}), we get

\begin{equation}
\label{2.10}
\Delta_{ab}+X_{ac}\,\omega^{cd}\,X_{bd}=0\,.
\end{equation}

\bigskip\noindent 
We notice that this equation does not give $X_{ab}$ univocally,
because it also contains the still unknown $\omega^{ab}$. What we
usually do is to choose $\omega^{ab}$ in such a way that the new
variables are unconstrained. It is opportune to mention that it is
not always possible to make a choice like this, that is to say,
just having unconstrained variables \cite{Barc}. In consequence, the
consistency of the method requires an introduction of other new
variables in order to transform these constraints also into
first-class. This may lead to an endless process. However, it is
important to emphasize that $\omega^{ab}$ can be fixed anyway.

\medskip
After fixing $\omega^{ab}$, we pass to consider the coefficients
$X_{ab}$. They are also not obtained in a univocally way. Even after
fixing $\omega^{ab}$, expression (\ref{2.10}) leads to less equations
than variables. The choice of $X$'s has to be done in a convenient
way \cite{Banerjee}. We shall see in the following sections how to
use this freedom in order to keep the supersymmetry in the final
theory. 

\medskip
The knowledge of $X_{ab}$ permits us to obtain $T_a^{(1)}$. If
$X_{ab}$ do not depend on $(q,p)$, it is easily seen that
$T_a+T_a^{(1)}$ is already strongly involutive. When this occurs, we
succeed in obtaining $\tilde T_a$. If this is not so, we have to
introduce $T_a^{(1)}$ into equation~(\ref{2.8b}) to calculate
$T_a^{(2)}$, and so on.

\medskip
Another point in the BFFT formalism is that any dynamic function
$A(q,p)$ (for instance, the Hamiltonian) has also to be properly
modified in order to be strongly involutive with the first-class
constraints $\tilde T_a$. Denoting the modified quantity by $\tilde
A(q,p;y)$, we then have

\begin{equation}
\label{2.11}
\bigl\{\tilde T_a,\,\tilde A\bigr\}=0\,.
\end{equation}

\bigskip\noindent
In addition, $\tilde A$ has also to satisfy  the boundary condition

\begin{equation}
\label{2.12}
\tilde A(q,p;0)=A(q,p)\,.
\end{equation}

\bigskip
The obtainment of $\tilde A$ is similar to what was done to get
$\tilde T_a$, that is to say, we consider an expansion like

\begin{equation}
\label{2.13}
\tilde A=\sum_{n=0}^\infty A^{(n)}\,,
\end{equation}

\bigskip\noindent 
where $A^{(n)}$ is also a term of order $n$ in $y$'s.
Consequently, compatibility with~(\ref{2.12}) requires that

\begin{equation}
\label{2.14}
A^{(0)}=A\,.
\end{equation}

\bigskip
\noindent The combination of~(\ref{2.6}),~(\ref{2.11})
and~(\ref{2.13}) gives

\begin{eqnarray}
\bigl\{T_a^{(0)},A^{(0)}\bigr\}_{(q,p)}
&+&\bigl\{T_a^{(1)},A^{(1)}\bigr\}_{(y)}=0\,,
\label{2.15a}\\
\bigl\{T_a^{(0)},A^{(1)}\bigr\}_{(q,p)}
&\!\!\!\!\!+\!\!\!\!\!&\bigl\{T_a^{(1)},A^{(0)}\bigr\}_{(q,p)}
+\bigl\{T_a^{(1)},A^{(2)}\bigr\}_{(y)}
+\bigl\{T_a^{(2)},A^{(1)}\bigr\}_{(y)}=0\,,
\label{2.15b}\\
\bigl\{T_a^{(0)},A^{(2)}\bigr\}_{(q,p)}
&\!\!\!\!\!+\!\!\!\!\!&\bigl\{T_a^{(1)},A^{(1)}\bigr\}_{(q,p)}
+\bigl\{T_a^{(2)},A^{(0)}\bigr\}_{(q,p)}
+\bigl\{T_a^{(1)},A^{(3)}\bigr\}_{(y)}
\nonumber\\
&\!\!\!\!\!+\!\!\!\!\!&\bigl\{T_a^{(2)},A^{(2)}\bigr\}_{(y)}
+\bigl\{T_a^{(3)},A^{(1)}\bigr\}_{(y)}=0\,,
\label{2.15c}\\
&\vdots&
\nonumber
\end{eqnarray}

\bigskip\noindent
which correspond to the coefficients of the powers $y^0$,
$y^1$, $y^2$, etc., respectively. The expression~(\ref{2.15a})
above gives us $A^{(1)}$

\begin{equation}
\label{2.16}
A^{(1)}=-\,y^a\,\omega_{ab}\,X^{bc}\,
\bigl\{T_c,\,A\bigr\}\,,
\end{equation}

\bigskip\noindent 
where $\omega_{ab}$ and $X^{ab}$ are the inverses of $\omega^{ab}$
and $X_{ab}$.

\medskip
In the obtainment of $T^{(1)}_a$ we had seen that $T_a+T^{(1)}_a$ was
strongly involutive if the coefficients $X_{ab}$ do not depend on
$(q,p)$. However, the same argument does not necessarily apply here.
Usually we have to calculate other corrections to obtain the
final $\tilde A$. Let us discuss how this can be systematically done.
We consider the general case first.  The correction $A^{(2)}$ comes
from equation~(\ref{2.15b}), that we conveniently rewrite
as

\begin{equation}
\label{2.17}
\bigl\{T_a^{(1)},\,A^{(2)}\bigr\}_{(y)}
=-\,G_a^{(1)}\,,
\end{equation}

\bigskip\noindent
where

\begin{equation}
\label{2.18}
G_a^{(1)}=\bigl\{T_a,\,A^{(1)}\bigr\}_{(q,p)}
+\bigl\{T_a^{(1)},\,A\bigr\}_{(q,p)}
+\bigl\{T_a^{(2)},\,A^{(1)}\bigr\}_{(y)}\,.
\end{equation}

\bigskip\noindent 
Thus

\begin{equation}
\label{2.19}
A^{(2)}=-{1\over2}\,y^a\,\omega_{ab}\,X^{bc}\,G_c^{(1)}\,.
\end{equation}

\bigskip\noindent
In the same way, other terms can be obtained. The final general
expression reads

\begin{equation}
\label{2.20}
A^{(n+1)}=-{1\over n+1}\,y^a\,\omega_{ab}\,X^{bc}\,G_c^{(n)}\,,
\end{equation}

\bigskip\noindent 
where

\begin{equation}
\label{2.21}
G_a^{(n)}=\sum_{m=0}^n\bigl\{T_a^{(n-m)},\,A^{(m)}\bigr\}_{(q,p)}
+\sum_{m=0}^{n-2}\bigl\{T_a^{(n-m)},\,A^{(m+2)}\bigr\}_{(y)}
+\bigl\{T_a^{(n+1)},\,A^{(1)}\bigr\}_{(y)}\,.
\end{equation}

\bigskip\noindent
For the particular case when $X_{ab}$ do not depend on $(q,p)$ we
have that the corrections $A^{(n+1)}$ can be obtained by the same
expression~(\ref{2.20}), but $G_a^{(n)}$ simplifies to

\begin{equation}
\label{2.22}
G_a^{(n)}=\bigl\{T_a,\,A^{(n)}\bigr\}_{(q,p)}\,.
\end{equation}

\bigskip
At this stage, it is opportune to mention that when $X_{ab}$ do not
depend on $(q,p)$ and the second-class constraints are linear in the
phase-space variables, there is an alternative and more direct way of
calculating the modified function $\tilde A$~\cite{Das}. Let us
briefly discuss how this can be done. The combination of (\ref{2.20})
and (\ref{2.22}) gives

\begin{eqnarray}
A^{(n+1)}&=&-\,\frac{1}{n+1}\,y^a\omega_{ab}X^{bc}\,
\bigl\{T_c,\,A^{(n)}\bigr\}_{(q,p)}\,,
\nonumber\\
&=&-\frac{1}{n+1}\,y^a\omega_{ab}X^{bc}\,
\bigl\{T_c,\,\Omega^\alpha\bigr\}\,
\frac{\partial A^{(n)}}{\partial\Omega^\alpha}\,,
\nonumber\\
&\equiv&-\,\frac{1}{n+1}\,y^a\,K_a^\alpha\,
\partial_\alpha A^{(n)}\,,
\label{2.22a}
\end{eqnarray}

\bigskip\noindent
where $\Omega^\alpha$ is generically representing the phase-space
coordinates $(q,p)$ and

\begin{equation}
K_a^\alpha\equiv\omega_{ab}X^{bc}\,\bigr\{T_c,\,\Omega^\alpha\bigr\}
\label{2.22b}
\end{equation}

\bigskip\noindent 
is a constant matrix. By using expression~(\ref{2.22a}) iteratively,
we get 

\begin{equation}
A^{(n)}=\frac{(-1)^n}{n!}\,
\Bigl(y^a\,K_a^\alpha\,\partial_\alpha\Bigr)^n\,A\,.
\label{2.22c}
\end{equation}

\bigskip\noindent
Introducing (\ref{2.22c}) into (\ref{2.13}), we obtain $\tilde A$ in
terms of $A$, namely

\begin{eqnarray}
\tilde A(\Omega,y)&=&\sum_{n=0}^\infty\frac{(-1)^n}{n!}\,
\Bigl(y^a\,K_a^\alpha\,\partial_\alpha\Bigr)^n\,A\,,
\nonumber\\
&=&\exp\,\Bigl(-y^a\,K_a^\alpha\,\partial_\alpha\Bigr)\,A\,.
\label{2.22d}
\end{eqnarray}

\bigskip\noindent
Since $\exp\,(-y^a K_a^\alpha\partial_\alpha)$ is a translation
operator in the coordinate $\Omega$, we have

\begin{equation}
\tilde A\,(\Omega^\alpha,y^a)=A\,(\Omega^\alpha-y^aK_a^\alpha)
\label{2.22e}
\end{equation}

\bigskip\noindent
So, when $X_{ab}$ do not depend on the phase-space coordinates
$y^\alpha$ and the constraints $T_a$ are linear in $\Omega^\alpha$,
the dynamic quantity $\tilde A$ can be directly obtained from $A$ by
replacing $\Omega^\alpha$ by a shifted coordinate

\begin{equation}
\tilde \Omega^\alpha=\Omega^\alpha-y^a\,K_a^\alpha\,.
\label{2.22f}
\end{equation}

\bigskip
To conclude this brief report on the BFFT formalism, we refer to the
case where there are also first-class constrains in the theory. Let
us call them $F_\alpha$. We consider that the constraints of the
theory satisfy the following involutive algebra (with the use of the
Dirac bracket definition to strongly eliminate the second-class
constraints)

\begin{eqnarray}
\bigl\{F_\alpha,\,F_\beta\bigr\}_D
&=&U_{\alpha\beta}^\gamma\,F_\gamma
+I_{\alpha\beta}^a\,T_a\,,
\nonumber\\
\bigl\{H_0,\,F_\alpha\bigr\}_D
&=&V_\alpha^\beta\,F_\beta+K_\alpha^a\,T_a\,,
\nonumber\\
\bigl\{F_\alpha,\,T_b\bigr\}_D
&=&0
\label{2.23}
\end{eqnarray}

\bigskip\noindent
In the expressions above, $U_{\alpha\beta}^\gamma$,
$I_{\alpha\beta}^a$, $V_\alpha^\beta$ and $K_\alpha^a$ are structure
functions of the involutive algebra.

\medskip
The BFFT procedure in this case also introduces one auxiliary variable
for each one of the second-class constraints (this is also in
agreement with the counting of the physical degrees of freedom of the
initial theory). All the constraints and the Hamiltonian have to be
properly modified in order to satisfy the same involutive algebra
above, namely,

\begin{eqnarray}
\bigl\{\tilde F_\alpha,\,\tilde F_\beta\bigr\}
&=&U_{\alpha\beta}^\gamma\,\tilde F_\gamma
+I_{\alpha\beta}^a\,\tilde T_a\,,
\nonumber\\
\bigl\{\tilde H_0,\,\tilde F_\alpha\bigr\}
&=&V_\alpha^\beta\,\tilde F_\beta
+K_\alpha^a\,\tilde T_a\,,
\nonumber\\
\bigl\{\tilde F_\alpha,\,\tilde T_b\bigr\}
&=&0\,.
\label{2.24}
\end{eqnarray}

\bigskip\noindent 
Since the algebra is now weakly involutive, the iterative calculation
of the previous case cannot be applied here. We have to figure out
the corrections that have to be done in the initial quantities. For
details, see ref. \cite{BT}.

\vspace{1cm}
\section{An example in N=1 supersymmetry}
\renewcommand{\theequation}{3.\arabic{equation}}
\setcounter{equation}{0}

\bigskip
Let us discuss here a simple model developed on a parameter space
with just one Grassmannian variable. The supersymmetry
transformations are defined by

\begin{eqnarray}
\delta t&=&i\epsilon\,\theta\,,
\nonumber\\
\delta\theta&=&\epsilon\,,
\label{3.1}
\end{eqnarray}

\bigskip
\noindent
where $\epsilon$ is a constant anticommuting parameter that
characterizes the supersymmetry transformations (in this section we
consider both $\theta$ and $\epsilon$ as real quantities).

\medskip
The transformations above can be expressed in terms of a
supersymmetric $Q$-generator as follows

\begin{eqnarray}
\delta t&=&\bigl(\epsilon Q\bigr)\,t\,,
\nonumber\\
\delta\theta&=&\bigl(\epsilon Q\bigr)\,\theta\,.
\label{3.2}
\end{eqnarray}

\bigskip\noindent
The combination of (\ref{3.1}) and (\ref{3.2}) permit us to directly
infer that 

\begin{equation}
Q=\frac{\partial}{\partial\theta}
+i\theta\,\frac{\partial}{\partial t}
\label{3.3}
\end{equation}

\bigskip\noindent
and $Q$ satisfies the anticommuting algebra

\begin{equation}
\bigl[Q,Q\bigr]_+=2i\,\frac{\partial}{\partial t}\,.
\label{3.4}
\end{equation}

\bigskip
In this space, a (real) supercoordinate has the general form 

\begin{equation}
\phi^i(t,\theta)=q^i(t)+i\theta\,\psi^i(t)\,,
\label{3.5}
\end{equation}

\bigskip\noindent
where $q^i$ are bosonic coordinates and $\psi^i$ are fermionic ones.
The index $i$ ($i=1,\dots,N$) corresponds to the dimension of the
configuration space. The quantity $\phi^i$ given by (\ref{3.5}) is
actually a supercoordinates if it satisfies the supersymmetry
transformation

\begin{equation}
\delta\phi^i=\bigl(\epsilon Q\bigr)\,\phi^i\,.
\label{3.6}
\end{equation}

\bigskip\noindent
The combination of (\ref{3.3}), (\ref{3.5}), and (\ref{3.6}) yields
the transformation laws for the component fields, namely

\begin{eqnarray}
\delta q^i&=&i\epsilon\,\psi^i\,,
\nonumber\\
\delta\psi^i&=&-\,\epsilon\,\dot q^i\,.
\label{3.7}
\end{eqnarray}

\bigskip
Theories that are expressed in terms of supercoordinates and
operators that commute or anticommute with $Q$ are manifestly
supersymmetric invariant. An operator which anticommutes with $Q$ is 

\begin{equation}
D=\frac{\partial}{\partial\theta}
-i\theta\,\frac{\partial}{\partial t}\,.
\label{3.8}
\end{equation}

\bigskip\noindent
In order to construct supersymmetric Lagrangians, it is necessary to
introduce another operator. Here, this is nothing other than the time
derivative. For questions of hermicity, we conveniently write

\begin{equation}
D_t=i\,\frac{\partial}{\partial t}\,.
\label{3.9}
\end{equation}

\bigskip
We then consider the Lagrangian

\begin{equation}
{\cal L}=\frac{1}{2}\,D_t\phi^i\,D\phi^i\,,
\label{3.10}
\end{equation}

\bigskip\noindent
that in terms of components gives

\begin{equation}
{\cal L}=-\,\frac{1}{2}\,\dot q^i\psi^i
+\frac{1}{2}\,\theta\,
\bigl(\dot q^i\dot q^i-i\dot\psi^i\psi^i\bigr)\,.
\label{3.11}
\end{equation}

\bigskip\noindent
The $\theta$-component of the Lagrangian above transforms as a total
derivative. So, it is considered to be the Lagrangian in terms of the
component coordinate, namely

\begin{equation}
L=\frac{1}{2}\,\dot q^i\dot q^i
-\frac{i}{2}\,\dot\psi^i\psi^i\,.
\label{3.12}
\end{equation}

\bigskip\noindent
In order to identify the constraints, we calculate the canonical
momenta  

\begin{eqnarray}
p^i&=&\frac{\partial L}{\partial\dot q^i}=\dot q^i\,,
\nonumber\\
\pi^i&=&\frac{\partial L}{\partial\dot\psi^i}
=-\,\frac{i}{2}\,\psi^i\,,
\label{3.13}
\end{eqnarray}

\bigskip\noindent
where we have used left derivative for fermionic coordinates.  We
notice that the second relation above is a constraint, that we write
as 

\begin{equation}
T^i=\pi^i+\frac{i}{2}\,\psi^i\approx0\,,
\label{3.14}
\end{equation}

\bigskip\noindent
where $\approx$ means weakly equal \cite{Dirac}. This is the only
constraint of the theory and it is second-class. Its Poisson bracket
relation reads

\begin{eqnarray}
\Delta&=&\bigl\{T^i,\,T^j\bigr\}\,,
\nonumber\\
&=&-\,i\,\delta^{ij}\,.
\label{3.15}
\end{eqnarray}

\bigskip\noindent
The implementation of the BFFT formalism requires the introduction of
auxiliary variables $\xi^i$ (one for each second-class constraint
$T^i$). This coordinate is fermionic because the corresponding
constraint is also fermionic. Since there is just one kind of
auxiliary coordinates, we have to assume that they are constrained.
This is so in order to obtain strong first-class constraints. We then
consider

\begin{equation}
\omega^{ij}=\bigl\{\xi^i,\xi^j\bigr\}=-\,i\,\delta^{ij}\,.
\label{3.16}
\end{equation}

\bigskip\noindent
For this particular example, we do not need to go into details in the
BFFT formalism to conclude that the first class constraints $\tilde
T^i$ are given by

\begin{equation}
\tilde T^i=\pi^i+\frac{i}{2}\,\psi^i+i\,\xi^i\approx0\,.
\label{3.17}
\end{equation}

\bigskip\noindent
Further, it is also immediately seen that the Lagrangian that leads
to this theory reads

\begin{equation}
L=\frac{1}{2}\,\dot q^i\dot q^i
-\,\frac{i}{2}\,\dot\psi^i\psi^i
-\,i\,\dot\xi^i\xi^i
+\,i\,\xi^i\dot\psi^i\,.
\label{3.18}
\end{equation}

\bigskip\noindent
We observe that it will be supersymmetric invariant in the sense of
the following transformations 

\begin{eqnarray}
\delta q^i&=&i\,\epsilon\,\bigl(\psi^i+2\xi^i\bigr)\,,
\nonumber\\
\delta\psi^i&=&\epsilon\,\dot q^i\,,
\nonumber\\
\delta\xi^i&=&-\,\epsilon\,\dot q^i\,.
\label{3.19}
\end{eqnarray}

\bigskip
This is a very simple model and of course cannot tell us everything
about the BFFT formalism for supersymmetric theories. But it is
important to emphasize that the obtained effective theory is also
constrained in the auxiliary variables. If we would like to go on
with the BFFT method, we would have to introduce new auxiliary
variables to take care of these new constraints. It is easily seen
that these new variables would also have to be constrained and that
this procedure would lead to an endless process, with an infinite
number of auxiliary variables. We shall see that this characteristic
will also be manifested in the example we shall discuss in the next
section.

\vspace{1cm}
\section{An example with N=2 supersymmetry}
\renewcommand{\theequation}{4.\arabic{equation}}
\setcounter{equation}{0}

\bigskip
Let us here consider the same supersymmetric quantum mechanical
system discussed in ref.~\cite{Witten}. It is developed in a space
containing two Grassmannian variables, $\theta$ and $\bar\theta$. The
supersymmetry transformations are defined by

\begin{eqnarray}
\delta t&=&i\bar\epsilon\,\theta+i\epsilon\,\bar\theta\,,
\nonumber\\
\delta\theta&=&\epsilon\,,
\nonumber\\
\delta\bar\theta&=&\bar\epsilon\,,
\label{4.1}
\end{eqnarray}

\bigskip\noindent 
where in this section we use the following convention for complex
conjugation: $\theta^\ast=\bar\theta$ and
$\epsilon^\ast=\bar\epsilon$. There are two supersymmetric
generators, $Q$ and $\bar Q$, defined by

\begin{eqnarray}
\delta t&=&\bigl(\epsilon Q+\bar\epsilon\bar Q\bigr)\,t\,,
\nonumber\\
\delta\theta&=&\bigl(\epsilon Q\bigr)\,\theta\,,
\nonumber\\
\delta\bar\theta&=&\bigl(\bar\epsilon\bar Q\bigr)\,\bar\theta\,.
\label{4.2}
\end{eqnarray}

\bigskip\noindent
From (\ref{4.1}) and (\ref{4.2}) one can directly infer

\begin{eqnarray}
Q&=&\frac{\partial}{\partial\theta}
+i\bar\theta\,\frac{\partial}{\partial t}\,,
\nonumber\\
\bar Q&=&\frac{\partial}{\partial\bar\theta}
+i\theta\,\frac{\partial}{\partial t}\,.
\label{4.3}
\end{eqnarray}

\bigskip\noindent
These operators satisfy the algebra

\begin{eqnarray}
&&\bigl[Q,\bar Q\bigr]_+=2i\,\frac{\partial}{\partial t}\,,
\nonumber\\
&&\bigl[Q,Q\bigr]_+=0=\bigl[\bar Q,\bar Q\bigr]_+\,.
\label{4.4}
\end{eqnarray}

\bigskip\noindent
Here, the (real) supercoordinate acquires the form

\begin{equation}
\phi^i(t,\theta,\bar\theta)=q^i(t)
+i\theta\,\psi^i(t)+i\bar\theta\,\bar\psi^i(t)
+\bar\theta\theta\,d^i(t)\,,
\label{4.5}
\end{equation}

\bigskip\noindent
that must satisfy the supersymmetry transformation

\begin{equation}
\delta\phi^i=\bigl(\epsilon Q+\bar\epsilon\bar Q\bigr)\,\phi^i\,.
\label{4.6}
\end{equation}

\bigskip\noindent
The coordinates $\psi^i$ and $\bar\psi^i$ have the same complex
conjugation as $\theta$ and $\bar\theta$. The combination of
(\ref{4.3}), (\ref{4.5}) and (\ref{4.6}) gives the following
transformation for the component coordinates:

\begin{eqnarray}
\delta q^i&=&i\,\bigl(\epsilon\psi^i+\bar\epsilon\bar\psi^i\bigr)\,,
\nonumber\\
\delta\psi^i&=&-\,\bar\epsilon\,\bigl(\dot q^i-i\,d^i\bigr)\,,
\nonumber\\
\delta\bar\psi^i&=&-\,\epsilon\,\bigl(\dot q^i+i\,d^i\bigr)\,,
\nonumber\\
\delta d^i&=&\bar\epsilon\dot{\bar\psi^i}-\epsilon\dot\psi^i\,.
\label{4.7}
\end{eqnarray}

\bigskip\noindent
There are two operators that anticommute with $Q$ and $\bar Q$, which
are

\begin{eqnarray}
D&=&\frac{\partial}{\partial\theta}
-i\bar\theta\,\frac{\partial}{\partial t}\,,
\nonumber\\
\bar D&=&\frac{\partial}{\partial\bar\theta}
-i\theta\,\frac{\partial}{\partial t}\,.
\label{4.8}
\end{eqnarray}

\bigskip\noindent
With these operators and the supercoordinates we may construct the
following general Lagrangian density

\begin{equation}
{\cal L}=\frac{1}{2}\,\bar D\phi^i\,D\phi^i-V(\phi)\,.
\label{4.9}
\end{equation}

\bigskip\noindent
The component Lagrangian is obtained by taking the $\bar\theta\theta$
part of the Lagrangian above. The result is

\begin{equation}
L=\frac{1}{2}\,\dot q^i\dot q^i
+i\,\bar\psi^i\dot\psi^i
+\frac{1}{2}\,d^id^i
-d^i\,\frac{\partial v}{\partial q^i}
+\psi^i\bar\psi^j\,\frac{\partial^2v}{\partial q^i\partial q^j}\,,
\label{4.10}
\end{equation}

\bigskip\noindent
where $v=V(q)$. The canonical momenta conjugate to $q^i$, $d^i$,
$\psi^i$ and $\bar\psi^i$ are given respectively by

\begin{eqnarray}
p^i&=&\frac{\partial L}{\partial\dot q^i}=\dot q^i\,,
\nonumber\\
s^i&=&\frac{\partial L}{\partial\dot d^i}=0\,,
\nonumber\\
\pi^i&=&\frac{\partial L}{\partial\dot\psi^i}=-\,i\bar\psi^i\,,
\nonumber\\
\bar\pi^i&=&\frac{\partial L}{\partial\dot{\bar\psi^i}}=0\,.
\label{4.11}
\end{eqnarray}

\bigskip\noindent
From these equations, one identifies the constraints

\begin{eqnarray}
T_1^i&=&\pi^i+i\,\bar\psi^i\approx0\,,
\nonumber\\
T_2^i&=&\bar\pi^i\approx0\,,
\nonumber\\
T_3^i&=&s^i\approx0\,.
\label{4.12}
\end{eqnarray}

\bigskip\noindent
Constructing the total Hamiltonian and imposing the consistency
condition that constraints do not evolve in time \cite{Dirac} we get
a new constraint

\begin{equation}
T_4^i=d^i-\frac{\partial v}{\partial q^i}\approx0\,,
\label{4.13}
\end{equation}

\bigskip\noindent
We observe that it is the equation of motion with respect the
coordinates $d^i$. We also mention that there are no more
constraints. 

\medskip
These all constraints are second-class. The matrix elements of their
Poisson brackets read

\begin{eqnarray}
\bigl(\Delta^{ij}\bigr)_{12}&=&-\,i\,\delta^{ij}\,,
\nonumber\\
&=&\bigl(\Delta^{ij}\bigr)_{21}\,,
\nonumber\\
\bigl(\Delta^{ij}\bigr)_{34}&=&-\,\delta^{ij}\,,
\nonumber\\
&=&-\,\bigl(\Delta^{ij}\bigr)_{43}\,.
\label{4.14}
\end{eqnarray}

\bigskip\noindent
Other elements are zero.

\medskip
To implement the BFFT formalism, we introduce auxiliary coordinates,
one for each second-class constraint. Let us generically denote them
by $y^{ai}$, where $a=1,2,3,4$. These are anticommuting coordinates
for $a=1,2$ and commuting for $a=3,4$.

\medskip
Here, contrarily to the previous example, the number of constraints
is even. So, it is natural that we try to start by supposing that the
new variables are canonical and that there are no constrains
involving them. If we proceeded in this way, we would actually get a
final theory with just first-class constraints, but it would not be
supersymmetric. In fact, we mention that it would be completely
inconsistent (see appendix). Also here we have to consider that the
fermionic coordinates are not canonical (but there is no problem
concerning the bosonic sector) in order to get a final theory also
supersymmetric. Let us particularly choose

\begin{equation}
y^{ai}=\bigl(\xi^i,\,\bar\xi^i;\,\eta^i,\,\rho^i\bigr)\,.
\label{4.15}
\end{equation}

\bigskip\noindent
We shall consider the following bracket structure among these
variables 

\begin{eqnarray}
\bigl\{\xi^i,\bar\xi^j\bigr\}&=&-\,i\,\delta^{ij}\,,
\nonumber\\
\bigl\{\eta^i,\rho^j\bigr\}&=&\delta^{ij}\,.
\label{4.16}
\end{eqnarray}

\bigskip\noindent
Other brackets are zero. We notice that we have considered that
$\rho^i$ are the canonical momenta conjugate to $\eta^i$. With these
elements, the matrix $\omega$ defined at (\ref{2.2}) reads

\begin{equation}
\bigl[\bigl(\omega^{ij}\bigr)^{ab}\bigr]=\left(
\begin{array}{cccc}
0&-\,i&0&0\\
-\,i&0&0&0\\
0&0&0&1\\
0&0&-1&0
\end{array}
\right)\,\delta^{ij}\,,
\label{4.17}
\end{equation}

\bigskip\noindent
where rows and columns follow the order $\xi$, $\bar\xi$, $\eta$, and
$\rho$. 

\medskip 
Introducing this result into eq. (\ref{2.10}), as well as the ones
given by (\ref{4.14}), and considering that coefficients
$(X^{ij})_{ab}$ are bosonic for the combinations ($a=1,2$; $b=1,2$)
and ($a=3,4$; $b=3,4$), and fermionic for ($a=1,2$; $b=3,4$) and
($a=3,4$; $b=1,2$), we obtain the following independent equations

\begin{eqnarray}
i\delta^{ij}
+i\,\bigl(X^{ik}\bigr)_{11}\bigl(X^{jk}\bigr)_{22}
+i\,\bigl(X^{ik}\bigr)_{12}\bigl(X^{jk}\bigr)_{21}
&\!\!-\!\!&\,\bigl(X^{ik}\bigr)_{13}\bigl(X^{jk}\bigr)_{24}
\nonumber\\
&\!\!+\!\!&\,\bigl(X^{ik}\bigr)_{14}\bigl(X^{jk}\bigr)_{23}=0\,,
\nonumber\\
\delta^{ij}
+i\,\bigl(X^{ik}\bigr)_{31}\bigl(X^{jk}\bigr)_{42}
+i\,\bigl(X^{ik}\bigr)_{32}\bigl(X^{jk}\bigr)_{41}
&\!\!-\!\!&\,\bigl(X^{ik}\bigr)_{33}\bigl(X^{jk}\bigr)_{44}
\nonumber\\
&\!\!+\!\!&\,\bigl(X^{ik}\bigr)_{34}\bigl(X^{jk}\bigr)_{43}=0\,,
\nonumber\\
i\,\bigl(X^{ik}\bigr)_{11}\bigl(X^{jk}\bigr)_{32}
+i\,\bigl(X^{ik}\bigr)_{12}\bigl(X^{jk}\bigr)_{31}
&\!\!-\!\!&\,\bigl(X^{ik}\bigr)_{13}\bigl(X^{jk}\bigr)_{34}
\nonumber\\
&\!\!+\!\!&\,\bigl(X^{ik}\bigr)_{14}\bigl(X^{jk}\bigr)_{33}=0\,,
\nonumber\\
i\,\bigl(X^{ik}\bigr)_{11}\bigl(X^{jk}\bigr)_{42}
+i\,\bigl(X^{ik}\bigr)_{12}\bigl(X^{jk}\bigr)_{41}
&\!\!-\!\!&\,\bigl(X^{ik}\bigr)_{13}\bigl(X^{jk}\bigr)_{44}
\nonumber\\
&\!\!+\!\!&\,\bigl(X^{ik}\bigr)_{14}\bigl(X^{jk}\bigr)_{43}=0\,,
\nonumber\\
i\,\bigl(X^{ik}\bigr)_{21}\bigl(X^{jk}\bigr)_{32}
+i\,\bigl(X^{ik}\bigr)_{22}\bigl(X^{jk}\bigr)_{31}
&\!\!-\!\!&\,\bigl(X^{ik}\bigr)_{23}\bigl(X^{jk}\bigr)_{34}
\nonumber\\
&\!\!+\!\!&\,\bigl(X^{ik}\bigr)_{24}\bigl(X^{jk}\bigr)_{33}=0\,,
\nonumber\\
i\,\bigl(X^{ik}\bigr)_{21}\bigl(X^{jk}\bigr)_{42}
+i\,\bigl(X^{ik}\bigr)_{22}\bigl(X^{jk}\bigr)_{41}
&\!\!-\!\!&\,\bigl(X^{ik}\bigr)_{23}\bigl(X^{jk}\bigr)_{44}
\nonumber\\
&\!\!+\!\!&\,\bigl(X^{ik}\bigr)_{24}\bigl(X^{jk}\bigr)_{43}=0\,.
\label{4.18}
\end{eqnarray}

\bigskip
The set of equations above does not univocally fix $(X^{ij})_{ab}$.
There are less equations than variables. We are then free to choose
some of these coefficients in a convenient way. First, we notice that
it is a good simplification to consider them bosonic. We thus take
equal to zero the fermionic ones, i.e.

\begin{equation}
\bigl(X^{ij}\bigr)_{ab}=0
\hspace{1cm}{\rm for}\,\,\,(a=1,2;b=3,4)
\,\,\,{\rm and}\,\,\,(a=3,4;b=1,2)\,.
\label{4.24}
\end{equation}

\bigskip\noindent
With this choice, the set of equations above simplifies to

\begin{eqnarray}
\delta^{ij}
+\bigl(X^{ik}\bigr)_{11}\bigl(X^{jk})_{22}
+\bigl(X^{ik}\bigr)_{12}\bigl(X^{jk})_{21}&=&0\,,
\nonumber\\
\delta^{ij}
-\bigl(X^{ik}\bigr)_{33}\bigl(X^{jk})_{44}
+\bigl(X^{ik}\bigr)_{34}\bigl(X^{jk})_{43}&=&0\,.
\label{4.26}
\end{eqnarray}

\bigskip\noindent
Looking at the expressions of the constraints $T_a^i$, we notice que
a choice that might preserve the supersymmetry is

\begin{eqnarray}
\bigl(X^{ij}\bigr)_{12}&=&i\,\delta^{ij}\,,
\nonumber\\
\bigl(X^{ij}\bigr)_{21}&=&i\,\delta^{ij}\,,
\nonumber\\
\bigl(X^{ij}\bigr)_{11}&=&0=\bigl(X^{ij}\bigr)_{22}\,,
\nonumber\\
\bigl(X^{ij}\bigr)_{34}&=&-\,\delta^{ij}\,,
\nonumber\\
\bigl(X^{ij}\bigr)_{43}&=&\delta^{ij}\,,
\nonumber\\
\bigl(X^{ij}\bigr)_{33}&=&0=\bigl(X^{ij}\bigr)_{44}\,,
\label{4.27}
\end{eqnarray}

\bigskip\noindent
because this would lead to convenient translations in the old
coordinates.  Since it was possible to obtain the quantities $X$'s
independently of the phase space coordinates, we have that $T^{(1)}$,
given by (\ref{2.9}), is the entire contribution for $\tilde T$. Then
the set of new constraints reads

\begin{eqnarray}
\tilde T^i_1&=&\pi^i+i\,\bar\psi^i+i\,\bar\xi^i\,,
\nonumber\\
\tilde T^i_2&=&\bar\pi^i+i\,\xi^i\,,
\nonumber\\
\tilde T^i_3&=&s^i-\rho^i\,,
\nonumber\\
\tilde T^i_4&=&d^i-\frac{\partial v}{\partial q^i}+\eta^i\,.
\label{4.28}
\end{eqnarray}

\bigskip\noindent
Notice that the choice of the coefficients $X$'s were made such that
the auxiliary coordinates $\bar\xi^i$ and $\eta^i$ figure as
translations of $\bar\psi^i$ and $d^i$ respectively. This will make
easier our goal of obtaining a final supersymmetric theory.

\bigskip
The next step is to look for the Lagrangian that leads to this new
theory. A consistently way of doing this is by means of the path
integral formalism, where the Faddeev-Senjanovic procedure
\cite{Faddeev} has to be used. Let us then obtain the canonical
Hamiltonian of the initial theory. We first write down the total
Hamiltonian 

\begin{eqnarray}
H=\dot q^ip^i+\dot d^is^i+\dot\psi^i\pi
+\dot{\bar\psi^i}\bar\pi^i-L+\lambda_aT_a\,.
\label{4.29}
\end{eqnarray}

\bigskip\noindent
The order of velocities and momenta (mainly for the Grassmannian
coordinates) is due to the left derivative convention. Using the
first equation (\ref{4.11}), one eliminates de velocities $\dot q^i$.
The remaining ones are eliminated by just redefining some Lagrange
multipliers. The final result reads

\begin{eqnarray}
H=\frac{1}{2}\,p^ip^i-\frac{1}{2}\,d^id^i
+d^i\,\frac{\partial v}{\partial q^i}
&\!\!\!+\!\!\!&\bar\psi^i\psi^j\,
\frac{\partial^2v}{\partial q^i\partial q^j}
+\bar\lambda_1^i\,\bigl(\pi^i+i\bar\psi^i\bigr)
\nonumber\\
&\!\!\!+\!\!\!&\bar\lambda_2^i\,\bar\pi^i
+\bar\lambda_3^i\,s^i
+\lambda_4^i\,\bigl(d^i-\frac{\partial v}{\partial q^i}\bigr)\,,
\label{4.30}
\end{eqnarray}

\bigskip\noindent
where $\bar\lambda$'s are the redefined Lagrange multipliers. From
the expression above one can identify the initial canonical
Hamiltonian

\begin{equation}
H_c=\frac{1}{2}\,p^ip^i-\frac{1}{2}\,d^id^i
+d^i\,\frac{\partial v}{\partial q^i}
+\bar\psi^i\psi^j\,\frac{\partial^2v}{\partial q^i\partial q^j}\,.
\label{4.31}
\end{equation}

\bigskip
To calculate the modified Hamiltonian we can use relations
(\ref{2.13}) and (\ref{2.20}), where $(\omega^{ij})_{ab}$ is the
inverse of (\ref{4.17}) and $(X^{ij})^{ab}$ is the inverse of the
matrix formed by relations (\ref{4.27}). On the other hand, it is
just a matter of algebraic calculation to obtain the quantities
$G_a^{(n)i}$, defined by (\ref{2.21}). The result is

\begin{eqnarray}
G_1^{(0)i}&=&\bar\psi^j\,
\frac{\partial^2v}{\partial q^i\partial q^j}\,,
\nonumber\\
G_2^{(0)i}&=&-\,\psi^j\,
\frac{\partial^2v}{\partial q^i\partial q^j}\,,
\nonumber\\
G_3^{(0)i}&=&d^i-\frac{\partial v}{\partial q^i}\,,
\nonumber\\
G_4^{(0)i}&=&-\,p^j\,
\frac{\partial^2v}{\partial q^i\partial q^j}\,,
\nonumber\\
G_1^{(1)i}&=&\bar\xi^j\,
\frac{\partial^2v}{\partial q^i\partial q^j}\,,
\nonumber\\
G_2^{(1)i}&=&-\,\xi^j\,
\frac{\partial^2v}{\partial q^i\partial q^j}\,,
\nonumber\\
G_3^{(1)i}&=&\eta^i\,,
\nonumber\\
G_4^{(1)i}&=&-\,\rho^j\,
\frac{\partial^2v}{\partial q^i\partial q^k}\,
\frac{\partial^2v}{\partial q^j\partial q^k}\,.
\label{4.32}
\end{eqnarray}

\bigskip\noindent
Using these quantities, we obtain the extended canonical Hamiltonian

\begin{eqnarray}
\tilde H_c&=&H_c+H_c^{(1)}+H_c^{(2)}
\nonumber\\
&=&\frac{1}{2}\,p^ip^i
-\frac{1}{2}\,d^id^i
-\frac{1}{2}\,\eta^i\eta^i-d^i\eta^i
+\bigl(d^i+\eta^i)\,\frac{\partial v}{\partial q^i}
\nonumber\\
&&\phantom{\frac{1}{2}\,p^ip^i}
+\Bigl(\bar\psi^i\psi^j+\bar\xi^i\xi^j
+\bar\xi^i\psi^j+\bar\psi^i\xi^j+\rho^ip^j\Bigr)\,
\frac{\partial^2v}{\partial q^i\partial q^j}
\nonumber\\
&&\phantom{\frac{1}{2}\,p^ip^i}
+\frac{1}{2}\,\rho^i\rho^j\,
\frac{\partial^2v}{\partial q^i\partial q^k}\,
\frac{\partial^2v}{\partial q^j\partial q^k}\,.
\label{4.33}
\end{eqnarray}

\bigskip
Since the quantities $(X^{ij})_{ab}$ do not depend on the initial
phase-space coordinates, we can check the result above by considering
the extended Hamiltonian (\ref{4.33}) could also have been obtained
from shifted coordinates defined in (\ref{2.22f}), i.e.,

\begin{equation}
\tilde H_c(\Omega,y)=H_c(\tilde\Omega)\,.
\label{4.34}
\end{equation}

\bigskip\noindent
The calculation of the shifted coordinates gives

\begin{eqnarray}
\tilde q^i&=&q^i\,,
\nonumber\\
\tilde p^i&=&p^i
+\rho^j\,\frac{\partial^2v}{\partial q^i\partial q^j}\,,
\nonumber\\ 
\tilde d^i&=&d^i+\eta^i\,,
\nonumber\\
\tilde s^i&=&s^i-\rho^i\,,
\nonumber\\
\tilde\psi^i&=&\psi^i+\xi^i\,,
\nonumber\\
\tilde{\bar\psi^i}&=&\bar\psi^i+\bar\xi^i\,,
\nonumber\\
\tilde\pi^i&=&\pi^i\,,
\nonumber\\
\tilde{\bar\pi}^i&=&\bar\pi^i+i\,\xi^i\,.
\label{4.35}
\end{eqnarray}

\bigskip\noindent
Notice once more that the choice we have made for the $X$'s
quantities at (\ref{4.27}) led to convenient shifts in the old
coordinates. We shall see that this will be very important to
demonstrate de supersymmetry invariance of the final theory. It is
just a matter of algebraic calculation to show that the combination
of (\ref{4.34}) and (\ref{4.35}) gives the extended canonical
Hamiltonian (\ref{4.33}).

\medskip
To use the path integral formalism to obtain the Lagrangian that
leads to this theory we have to pay attention for the fact that there
are constraints with respect the auxiliary variables $\xi^i$ and
$\bar\xi^i$. It is not difficult to see that a possible set of
constraints is

\begin{eqnarray}
\pi^i_\xi&=&-\,i\,\bar\xi^i\,,
\nonumber\\
\pi^i_{\bar\xi}&=&0\,.
\label{4.36}
\end{eqnarray}

\bigskip\noindent
The general expression for the propagator reads

\begin{equation}
Z=N\int[d\mu]\,\exp\,\Bigl\{i\int_{t_0}^tdt\,
\Bigl(\dot q^ip^i+\dot d^is^i+\dot\psi^i\pi^i
+\dot{\bar\psi^i}\bar\pi^i+\dot\eta^i\rho^i
+\dot\xi^i\pi^i_\xi+\dot{\bar\xi^i}\pi^i_{\bar\xi}
-\tilde H_c\Bigr)\Bigr\}\,,
\label{4.37}
\end{equation}

\bigskip\noindent
with the measure $[d\mu]$ given by

\begin{eqnarray}
[d\mu]&=&[dq][dp][dd][ds][d\psi][d\bar\psi][d\pi][d\bar\pi]
[d\xi][d\bar\xi][d\eta][d\rho][d\pi_\xi][d\pi_{\bar\xi}]
\nonumber\\
&&\phantom{[dq][dp][dd][ds]}
\vert\det\,\{,\}\vert\,\delta\,[\pi^i_\xi+i\bar\xi^i]\,
\delta\,[\pi^i_{\bar\xi}]\,\prod_a\delta\,[\tilde T_a]\,
\delta\,[\tilde\Lambda_a]\,,
\label{4.38}
\end{eqnarray}

\bigskip\noindent
where $\tilde\Lambda_a$ are gauge-fixing functions corresponding to
the first-class constraints $\tilde T_a$ and the term
$\vert\det\,\{,\}\vert$ represents the determinant of all constraints
of the theory, including the gauge-fixing ones. The quantity $N$ that
appears in (\ref{4.38}) is the usual normalization factor. Using the
delta functions, we can directly eliminate de momenta $\pi^i$,
$\bar\pi^i$, $s^i$, $\pi_\xi^i$, and $\pi^i_{\bar\xi}$. The result is

\begin{eqnarray}
Z&=&N\int[dq][dp][dd][d\psi][d\bar\psi]
[d\xi][d\bar\xi][d\eta][d\rho]\,
\vert\det\,\{,\}\vert\,\delta\,[\tilde T_4]\,
\prod_a\delta\,[\tilde\Lambda_a]
\nonumber\\
&&\phantom{\int[dq][dp]}
\exp\,\Bigl\{i\int_{t_0}^tdt\,\Bigl[\dot q^ip^i
+\bigl(\dot d^i+\dot\eta^i\bigr)\,\rho^i
-\frac{1}{2}\,p^ip^i+\frac{1}{2}\,d^id^i
+\frac{1}{2}\,\eta^i\eta^i+\eta^id^i
\nonumber\\
&&\phantom{\int[dq][dp]}
-\,i\,\bigl(\dot{\bar\psi^i}\psi^i+\dot{\bar\xi^i}\xi^i
+\dot{\bar\psi^i}\xi^i+\dot{\bar\xi^i}\psi^i\bigr)
\nonumber\\
&&\phantom{\int[dq][dp]}
-\,\Bigl(\bar\psi^i\psi^j+\bar\xi^i\psi^j
+\bar\psi^i\xi^j+\bar\xi^i\xi^j+\rho^ip^j\Bigr)\,
\frac{\partial^2v}{\partial q^i\partial q^j}
\nonumber\\
&&\phantom{\int[dq][dp]}
-\,\frac{1}{2}\,\rho^i\rho^j\,
\frac{\partial^2v}{\partial q^i\partial q^k}
\frac{\partial^2v}{\partial q^k\partial q^j}\Bigr]\Bigr\}\,.
\label{4.39}
\end{eqnarray}

\bigskip\noindent
Finally, integrating over $p^i$ and  $\rho^i$ we obtain 

\begin{eqnarray}
Z&=&N\int[dq][dd][d\psi][d\bar\psi]
[d\xi][d\bar\xi][d\eta]\,
\vert\det\,\{,\}\vert\,
\delta\,[\dot d^i+\dot\eta^i-\dot q^j\,
\frac{\partial^2v}{\partial q^i\partial q^j}\bigr]\,
\nonumber\\
&&\phantom{\int[dq][dd]}
\delta\,[\tilde T_4]\,\prod_a\delta\,[\tilde\Lambda_a]\,
\exp\,\Bigl\{i\int_{t_0}^tdt\,
\Bigl[\frac{1}{2}\,\dot q^i\dot q^i
+\frac{1}{2}\,d^id^i
+\frac{1}{2}\,\eta^i\eta^i+\eta^id^i
\nonumber\\
&&\phantom{\int[dq][dd]}
-\,\bigl(d^i+\eta^i\bigr)\,\frac{\partial v}{\partial q^i}
-\,i\,\bigl(\dot{\bar\psi^i}\psi^i+\dot{\bar\xi^i}\xi^i
+\dot{\bar\psi^i}\xi^i+\dot{\bar\xi^i}\psi^i\bigr)
\nonumber\\
&&\phantom{\int[dq][dd]}
-\,\Bigl(\bar\psi^i\psi^j+\bar\xi^i\psi^j
+\bar\psi^i\xi^j+\bar\xi^i\xi^j\Bigr)\,
\frac{\partial^2v}{\partial q^i\partial q^j}\Bigr]\Bigr\}\,,
\label{4.41}
\end{eqnarray}

\bigskip\noindent
where the new $\delta$-function above came from the integrations over
$\rho^i$. We notice that it is nothing other than the time derivative
of constraint $\tilde T_4$. It is then just a consistency condition
and does not represent any new restriction over the coordinates of
the theory.

\medskip
From the propagator (\ref{4.41}), we identify the extended Lagrangian

\begin{eqnarray}
\tilde L&=&\frac{1}{2}\,\dot q^i\dot q^i
+\frac{1}{2}\,d^id^i+\frac{1}{2}\,\eta^i\eta^i+\eta^id^i
-\,\Bigl(d^i+\eta^i\Bigr)\,\frac{\partial v}{\partial q^i}
\nonumber\\
&&\phantom{\frac{1}{2}\,\dot q^i\dot q^i}
-\,i\,\bigl(\dot{\bar\psi}^i\psi^i+\dot{\bar\xi}^i\xi^i
+\dot{\bar\psi}^i\xi^i+\dot{\bar\xi}^i\psi^i\bigr)
\nonumber\\
&&\phantom{\frac{1}{2}\,\dot q^i\dot q^i}
-\,\Bigl(\bar\psi^i\psi^j+\bar\psi^i\xi^j
+\bar\xi^i\psi^j+\bar\xi^i\xi^j\Bigr)\,
\frac{\partial^2v}{\partial q^i\partial q^j}
\label{4.42}
\end{eqnarray}

\bigskip\noindent
The Lagrangian above is supersymmetric invariant if we consider the
transformations 

\begin{eqnarray}
\delta q^i&=&i\,\bigl[\epsilon\,(\psi^i+\xi^i)
+\bar\epsilon\,(\bar\psi^i+\bar\xi^i)\bigr]\,,
\nonumber\\
\delta\psi^i&=&-\,\bar\epsilon\,\bigl(\dot q^i-id^i\bigr)\,,
\nonumber\\
\delta\bar\psi^i&=&-\,\epsilon\,\bigl(\dot q^i+id^i\bigr)\,,
\nonumber\\
\delta d^i&=&\bar\epsilon\,\dot{\bar\psi}^i
-\epsilon\,\dot\psi^i\,,
\nonumber\\
\delta\xi^i&=&i\,\bar\epsilon\,\eta^i\,,
\nonumber\\
\delta\bar\xi^i&=&-\,i\,\epsilon\,\eta^i\,,
\nonumber\\
\delta \eta^i&=&\bar\epsilon\,\dot{\bar\xi}^i
-\epsilon\,\dot\xi^i\,.
\label{4.43}
\end{eqnarray}

\vspace{1cm}
\section{Conclusion}

\bigskip
We have used the method developed by Batalin, Fradkin, Fradkina, and
Tyutin in order to quantize two supersymmetric quantum mechanical
models, by transforming second-class into first-class constraints. We
have shown that final supersymmetric theories are only achieved by
taking constrained auxiliary variables. This procedure leads to an
endless process of introducing auxiliary constrained variables and
transforming them into first-class ones. This appears to be a general
characteristic of the use of the BFFT method in supersymmetric
theories.

\medskip
Although we have just considered supersymmetric quantum mechanical
models, the same could be directly done for field theories. Of
course, the algebraic work would be much bigger, but these would give
no new information comparing what we have already found here.

\vfill
\newpage
\noindent {\bf Acknowledgment:} This work is supported in part by
Conselho Nacional de Desenvolvimento Cient\'{\i}fico e Tecnol\'ogico
- CNPq, Financiadora de Estudos e Projetos - FINEP and Funda\c{c}\~ao
Universit\'aria Jos\'e Bonif\'acio - FUJB (Brazilian Research
Agencies).

\vspace{1cm}
\appendix
\renewcommand{\theequation}{A.\arabic{equation}}
\setcounter{equation}{0}
\section*{Appendix}

\bigskip
Let us discuss in this appendix the result we would obtain if we had
considered the two auxiliary fermionic coordinates as non constrained
and taken one as the momentum conjugate of the other, say $\xi^i$ and
$\chi^i$, satisfying the Poisson bracket relation

\begin{equation}
\bigl\{\xi^i,\chi^j\bigr\}=-\,\delta^{ij}\,.
\label{A.1}
\end{equation}

\bigskip\noindent
The matrix $[(\omega^{ij})^{ab}]$ would be

\begin{equation}
\bigl[\bigl(\omega^{ij}\bigr)^{ab}\bigr]=\left(
\begin{array}{cccc}
0&-\,1&0&0\\
-\,1&0&0&0\\
0&0&0&1\\
0&0&-1&0
\end{array}
\right)\,\delta^{ij}\,.
\label{A.1a}
\end{equation}

\bigskip\noindent
Also considering the condition (\ref{4.24}) we would obtain the
following equations involving the coefficients $X$'s

\begin{eqnarray}
i\,\delta^{ij}
+\bigl(X^{ik}\bigr)_{11}\bigl(X^{jk}\bigr)_{22}
+\bigl(X^{ik}\bigr)_{12}\bigl(X^{jk}\bigr)_{21}&=&0\,,
\nonumber\\
\delta^{ij}
-\bigl(X^{ik}\bigr)_{33}\bigl(X^{jk}\bigr)_{44}
+\bigl(X^{ik}\bigr)_{34}\bigl(X^{jk}\bigr)_{43}&=&0\,,
\label{A.2}
\end{eqnarray}

\bigskip\noindent
and a  good choice for $X$'s could be (because it leads to
translations in the old coordinates)

\begin{eqnarray}
\bigl(X^{ij}\bigr)_{11}&=&i\,\delta^{ij}\,,
\nonumber\\
\bigl(X^{ij}\bigr)_{22}&=&-\,\delta^{ij}\,,
\nonumber\\
\bigl(X^{ij}\bigr)_{12}&=&0
\,\,\,=\,\,\,\bigl(X^{ij}\bigr)_{21}\,,
\nonumber\\
\bigl(X^{ij}\bigr)_{34}&=&-\,\delta^{ij}\,,
\nonumber\\
\bigl(X^{ij}\bigr)_{43}&=&\delta^{ij}\,,
\nonumber\\
\bigl(X^{ij}\bigr)_{33}&=&0
\,\,\,=\,\,\,\bigl(X^{ij}\bigr)_{44}\,.
\label{A.3}
\end{eqnarray}

\bigskip\noindent
The new constraints $\tilde T$ read

\begin{eqnarray}
\tilde T^i_1&=&\pi^i+i\,\bar\psi^i+i\,\xi^i\,,
\nonumber\\
\tilde T^i_2&=&\bar\pi^i-\chi^i\,,
\nonumber\\
\tilde T^i_3&=&s^i-\rho^i\,,
\nonumber\\
\tilde T^i_4&=&d^i-\frac{\partial v}{\partial q^i}+\eta^i\,.
\label{A.4}
\end{eqnarray}

\bigskip\noindent
We actually notice that we have chosen $X$'s in such a way that $\xi$
and $\eta$ appear as translations of the $\bar\psi$ and $d$,
respectively.

\bigskip
Following the same procedure as before, we obtain the extended
Hamiltonian 

\begin{eqnarray}
\tilde H_c&=&\frac{1}{2}\,p^ip^i-\frac{1}{2}\,d^id^i
-\frac{1}{2}\,\eta^i\eta^i-\eta^id^i
+\bigl(d^i+\eta^i\bigr)\,\frac{\partial v}{\partial q^i}
\nonumber\\
&&\phantom{\frac{1}{2}\,p^ip^i}
+\,\Bigl(\bar\psi^i\psi^j+i\,\bar\psi^i\chi^j
+\xi^i\psi^j+i\,\xi^i\chi^j+p^i\rho^j\Bigr)\,
\frac{\partial^2v}{\partial q^i\partial q^j}
\nonumber\\
&&\phantom{\frac{1}{2}\,p^ip^i}
+\,\frac{1}{2}\,\rho^i\rho^j\,
\frac{\partial^2v}{\partial q^i\partial q^k}\,
\frac{\partial^2v}{\partial q^k\partial q^j}\,.
\label{A.5}
\end{eqnarray}

\bigskip\noindent
Apparently, there is no problem with this Hamiltonian, but
considering the general expression of the propagator and using the
$\delta$ functional to eliminate the momenta $\pi$, $\bar\pi$, and
$s$, as well as performing the integration over $\chi$, we obtain

\begin{eqnarray}
Z&=&N\int[dq][dp][dd][d\psi][d\bar\psi][d\eta][d\rho][d\xi]\,
\vert\det\,\{,\}\vert\,\delta\,[\tilde T_4]\,
\prod_a\delta\,[\tilde\Lambda^a]
\nonumber\\
&&\phantom{N\int[dq][dp]}
\delta\,\Bigl[\dot{\bar\psi^i}+\dot\xi^i
-\,i\,\bigl(\bar\psi^j+\xi^j\bigr)\,
\frac{\partial^2v}{\partial a^i\partial q^j}\Bigr]\,
\exp\,\Bigl\{i\int_{t_0}^tdt\,
\Bigl[\dot q^ip^i+\dot d^i\rho^i+\dot\eta^i\rho^i
\nonumber\\
&&\phantom{N\int[dq][dp]}
+\,\frac{1}{2}\,d^id^i+\frac{1}{2}\,\eta^i\eta^i
+d^i\eta^i-\bigl(d^i+\eta^i\bigr)\,
\frac{\partial v}{\partial q^i}
\nonumber\\
&&\phantom{N\int[dq][dp]}
+i\,\psi^i\,\Bigl(\dot{\bar\psi^i}+\dot\xi^i
-\,i\,\bigl(\bar\psi^j+\xi^j\bigr)\,
\frac{\partial^2v}{\partial q^i\partial q^j}\Bigr)
\nonumber\\
&&\phantom{N\int[dq][dp]}
-\,\frac{1}{2}\,\Bigl(p^i+\rho^j\,
\frac{\partial^2v}{\partial a^i\partial q^j}\Bigr)\,
\Bigl(p^i+\rho^k\,
\frac{\partial^2v}{\partial a^i\partial q^k}\Bigr)\Bigr]\Bigr\}\,.
\label{A.6}
\end{eqnarray}

\bigskip
The $\delta$ function that appear into the expression above was
obtained after integrating over $\chi$. We notice that the constraint
imposed by this $\delta$ function is not a consistency condition for
other constraints, as occurred with the $\delta$-function that appears
in (\ref{4.42}). More than that, we also notice that it causes the
elimination of all terms with $\psi$ and $\bar\psi$ in the
propagator.  Consequently, this would lead to a final effective
Lagrangian without these variables, what is obviously a non
consistent result because it does not matches the initial theory when
the auxiliary fields are turned off. It is also important to
emphasize that this result is not related to the particular choice we
have made for the coefficients $X^\prime s$ given at (\ref{A.4}). We
mention that this kind of inconsistency would persist for any choice
we make for $X$'s in equations (\ref{A.2}).

\end{document}